\documentclass[8.5pt,twoside,twocolumn]{article}
\usepackage[super,sort&compress,comma]{natbib} 
\usepackage{mhchem}
\usepackage{times,mathptmx}
\usepackage{sectsty}
\usepackage{balance} 

\usepackage{graphicx} 
\usepackage{lastpage}
\usepackage{afterpage}
\usepackage[format=plain,justification=raggedright,singlelinecheck=false,font=small,labelfont=bf,labelsep=space]{caption} 
\usepackage{fancyhdr}
\usepackage{color}
\usepackage{here}
\begin{document}

\thispagestyle{plain}
\fancypagestyle{plain}{
\fancyhead[L]{\includegraphics[height=8pt]{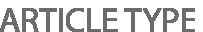}}
\fancyhead[C]{\hspace{-1cm}\includegraphics[height=20pt]{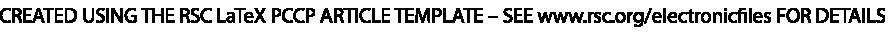}}
\fancyhead[R]{\includegraphics[height=10pt]{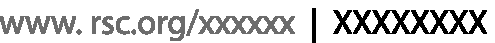}\vspace{-0.2cm}}
\renewcommand{\headrulewidth}{1pt}}
\renewcommand{\thefootnote}{\fnsymbol{footnote}}
\renewcommand\footnoterule{\vspace*{1pt}%
\hrule width 3.4in height 0.4pt \vspace*{5pt}} 
\setcounter{secnumdepth}{5}

\setcounter{topnumber}{2}
\setcounter{bottomnumber}{3}
\setcounter{totalnumber}{5}
\setcounter{dbltopnumber}{2}
\renewcommand\topfraction{.99}
\renewcommand\bottomfraction{.7}
\renewcommand\textfraction{.01}
\renewcommand\floatpagefraction{.6}
\renewcommand\dbltopfraction{.99}
\renewcommand\dblfloatpagefraction{.99}

\makeatletter 
\def\subsubsection{\@startsection{subsubsection}{3}{10pt}{-1.25ex plus -1ex minus -.1ex}{0ex plus 0ex}{\normalsize\bf}} 
\def\paragraph{\@startsection{paragraph}{4}{10pt}{-1.25ex plus -1ex minus -.1ex}{0ex plus 0ex}{\normalsize\textit}} 
\renewcommand\@biblabel[1]{#1}            
\renewcommand\@makefntext[1]%
{\noindent\makebox[0pt][r]{\@thefnmark\,}#1}
\makeatother 
\renewcommand{\figurename}{\small{Fig.}~}
\sectionfont{\large}
\subsectionfont{\normalsize} 

\fancyfoot{}
\fancyfoot[LO,RE]{\vspace{-7pt}\includegraphics[height=9pt]{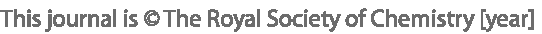}}
\fancyfoot[CO]{\vspace{-7.2pt}\hspace{12.2cm}\includegraphics{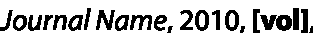}}
\fancyfoot[CE]{\vspace{-7.5pt}\hspace{-13.5cm}\includegraphics{RF}}
\fancyfoot[RO]{\footnotesize{\sffamily{1--\pageref{LastPage} ~\textbar  \hspace{2pt}\thepage}}}
\fancyfoot[LE]{\footnotesize{\sffamily{\thepage~\textbar\hspace{3.45cm} 1--\pageref{LastPage}}}}
\fancyhead{}
\renewcommand{\headrulewidth}{1pt} 
\renewcommand{\footrulewidth}{1pt}
\setlength{\arrayrulewidth}{1pt}
\setlength{\columnsep}{6.5mm}
\setlength\bibsep{1pt}

\twocolumn[
  \begin{@twocolumnfalse}
\noindent\LARGE{\textbf{Localization and size distribution of a polymer knot confined in a channel}
\vspace{0.6cm}

\noindent\large{\textbf{Chihiro H. Nakajima,\textit{$^{a\dag}$} Takahiro Sakaue,\textit{$^{ab\ddag}$} } }}\vspace{0.5cm}

\noindent\textit{\small{\textbf{Received Xth XXXXXXXXXX 20XX, Accepted Xth XXXXXXXXX 20XX\newline
First published on the web Xth XXXXXXXXXX 200X}}}

\noindent \textbf{\small{DOI: 10.1039/b000000x}}
\vspace{0.6cm}

\noindent \normalsize{
We have examined the behaviors of a knotted linear polymer in narrow tubes using Langevin
dynamics simulation to investigate the knot localization property in one-dimensional (1D) geometry.
We have found that the knot is strongly localized in such a geometry. 
By observing the distribution function of the size of localized knot, we found the scaling behavior of the fluctuation around the most probable size with radius of confinement.
Based on the analysis of the probability distribution of the knot size, we show that the strong localization behavior and the fluctuation around the most probable size can be encompassed by a simple argument based on the virtual tubes composed of parallel strands and overlapping among them.
}
\vspace{0.5cm}
 \end{@twocolumnfalse}
  ]

\section{Introduction}

\footnotetext{\dag~E-mail: nakajima@stat.phys.kyushu-u.ac.jp}
\footnotetext{\ddag~E-mail: sakaue@phys.kyusu-u.ac.jp}

\footnotetext{\textit{$^{a}$~Departent of Physics, Kyushu University, Fukuoka 812-8581, Japan}}
\footnotetext{\textit{$^{b}$~PRESTO, JST, Kawaguchi, Saitama 332-0012, Japan}}

Thanks to the advance in nanotechnology and single-molecule experiments, there has been a growing interest in the properties of biopolymers in confined spaces. 
For instance, it is now possible to bring genomic-length DNA molecules into $100$ nm size channels and to observe its static and dynamic behaviors\cite{TPCCRRWCSSH}.
For now, the existence of two different regimes depending on the channel size is well confirmed\cite{RFA}, and such a basic knowledge is expected to provide a guide for the controlling and manipulation of single biopolymers in fabricated devices. 

In the present paper, we look into statistical properties of intramolecular entanglements formed in a confined polymer in narrow channels.
Such topological defects are sometimes formed spontaneously in the course of experiments\cite{RB}, understanding of which would be of practical as well as fundamental importance.
In addition, they can be made by design using manipulation techniques\cite{Qu}, too.
We shall mainly address the question about the size and fluctuation of such defects.

The above question is directly related to the (de)localization phenomena of knots in closed ring polymers\cite{KOVDS,OW}.
It is known that the phenomenon of knot localization is strongly related to the geometry of the space in which a polymer can explore its large scale conformation.
Several recent studies have indicated that knots in three-dimensional (3D) space are likely to be weakly localized\cite{FKK,VKK,MOSZ,ST}.
In 2D space (realized by the confinement between parallel plates), the degree of the localization is enhanced, and the strong localization is expected for flat knots\cite{MHDKK,OSV}.
On the other hand, polymers assume a compact globular state either in poor solvent conditions or in small closed cavities, where knots are known to be delocalized~\cite{MOSZ,VKK}.

From this point of view, it is interesting to ask the behaviors of knots in 1D geometries, i.e., a chain confined in a narrow tube.
So far the study in this direction is not abundant.
In particular, there seems to be no investigation on the knot localization phenomena in 1D channels, except for a related work by M\"{o}bius et al.\cite{MFG}
They studied the unknotting dynamics of an initially knotted linear semiflexible polymer in 1D one channels, and predicted the pathway, in which the macroscopic inflation of knot takes place.
Here and in the followings, we use the term ``knot'' for the defect (intramolecular entanglement) even for open linear polymers, unless confusion is expected.

To clarify the static property of the knot in 1D confinement, we have numerically investigated the behavior of a linear chain containing a trefoil knot in a narrow tube\cite{NS}.
Although this trefoil may be eventually unknotted in the open chain with ends, one can naturally think of the precedent stage during which the knot exists in the identifiable manner.
Our results indicate that knots are strongly localized in 1D space and that the physical mechanism for the localization is entirely different from that in higher dimensions. In addition, the analysis of the knot
size distribution function reveals an important feature on the strand overlapping, which may be relevant to the statistics of ring polymers and star polymers confined in narrow tubes.


\section{Models and simulations}


\begin{figure}[h]
\centering
 \includegraphics[height=6cm]{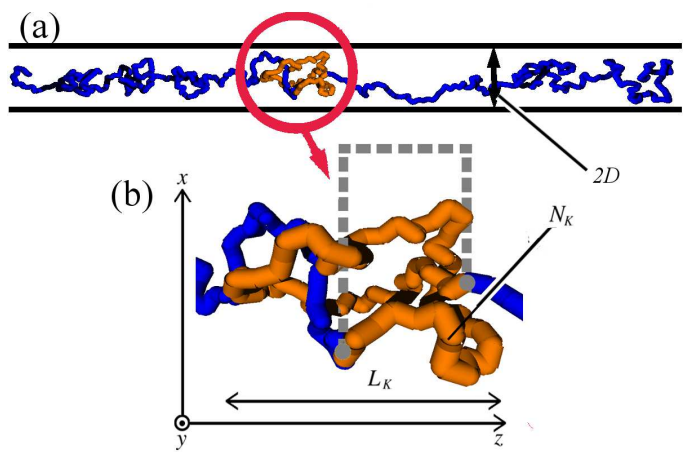}
  \caption{(Color online)Setup of the simulation. (a):A typical conformation which appeares in the simulation with $D=6$ and $N=500$. (b):Illustration for the closure scheme. The gray dashed line corresponds to the supplemental arc. The part with lighter (orange) color is detected as the knotted region.}
  \label{fgr:scheme}
\end{figure}

We have performed Langevin dynamics simulations to investigate the behavior of knotted polymers confined in a narrow tube.
We employ a linear chain model which consists of a sequence of beads of diameter $\sigma$ connected by springs:
\begin{eqnarray}\label{eq:bond}
U_{sp}&=&\frac{k_{sp}}{2\sigma^2}\sum_{i}\big( \ r_{i,i+1} -\sigma \ \big)^2,
\end{eqnarray}
where $r_{i,i+1}$ is the distance between neighboring beads, the large modulus $k_{sp}/k_BT=60$ ($k_BT$ being the thermal energy) keeps the bond length almost constant $\simeq \sigma$, and $\sigma$ is taken as a unity. 
A pair of beads interact by the repulsive Lennard-Jones potential:
\begin{eqnarray}\label{eq:int}
U_{int}&=&\left\{ \begin{array}{ll}
  \sum_{i \neq j } \ 4\epsilon \Big\{ \left(\frac{\sigma}{r_{i,j}}\right)^{12}-1 \Big\} & ( \ r_{i,j} \le \sigma \ ) \\
0 & ( \ r_{i,j} > \sigma \ )
\end{array} \right. ,
\end{eqnarray}
where $\epsilon = 0.8 k_BT $.
The polymer is confined in the cylindrical tube of radius $D$. 
By taking the tube axis as $z$ axis in the Cartesian coordinate, the wall potential is implemented as,
\begin{eqnarray}\label{eq:conf}
U_{W}&=&\left\{ \begin{array}{ll}
\sum_i \ 4\epsilon \Big\{ \left(\frac{\sigma}{D-r_{\perp i}}\right)^{12} -1 \Big\} & ( \ r_{i,j} \le \sigma \ ) \\
0 & ( \ r_{i,j} > \sigma \ )
\end{array} \right. ,
\end{eqnarray}
where $r_{\perp i}=\sqrt{x_i^2+y_i^2}$ is the distance of $i$-th bead from the tube axis.

The motion of a model polymer chain in thermal and viscous medium is described by the Langevin equation;
\begin{eqnarray}
m\frac{d^2\mathbf{r}_i}{dt^2}=-\gamma\frac{d\mathbf{r}_i}{dt}-\frac{\partial U}{\partial \mathbf{r}_i}+\mathbf{R}_i(t),
\end{eqnarray}
where $m$, $\gamma$ are mass and friction constant of monomers, $U=U_{sp}+U_{int}+U_{W}$ is the potnetial, and $\mathbf{R}_i(t)$ is a Gaussian white noise on the $i$-th monomer, which satisfies the following properies,
\begin{eqnarray}
\langle \mathbf{R}_i(t) \rangle=0 , \ \langle \mathbf{R}_i(t) \mathbf{R}_i(t')\rangle = 6\gamma k_B T\delta_{i,j}\delta(t-t').
\end{eqnarray}

As an initial configuration, a trefoil loop is introduced by hand near the middle of chain in an arbitrary way.
Starting from such a designed configuration, we equilibrate the system by numerically integrating the Langevin equation. Configurations were sampled after sufficiently long relaxation for the statistical analysis.
Identification of the knotted component is based on calculations of the special value of the Alexander polynomial; $\Delta (a)$ with $a=-1$.
Since the Alexander polynomial is defined for a closed ring, we need a closure scheme to apply it to our linear chain.
We label each monomer as $n=1,2, \cdots, N$ from one end, and make a closed ring by bridging $n_1$-th and $n_2$-th monomers with a supplemental arc (Fig. \ref{fgr:scheme}(b) ).
Initially, both ends are bridged ($n_1=1, \ n_2=N$); then, the calculation of the Alexander polynomial for the knot diagram, which is obtained from the configuration of closed trajectory projected onto $XZ$ plane, yields $\Delta (-1)=3$, indicating that the trefoil knot is present in the range between $n_1$ and $n_2$.
We then increase (decrease) $n_1$ ($n_2$) one by one by cutting monomers from both ends, which is followed by the calculation of the Alexander polynomial at each step. 
This iterative process is repeated until the point $n_1=\tilde{n_1}$ and $n_2=\tilde{n_2}$, at which the value of the Alexander polynomial changes to $\Delta (-1)=1$. This indicates that the closed trajectory obtained by bridging $\tilde{n_1}$ and $\tilde{n_2}$ has a trivial topology.
Then the knotted size is detected as $N_k = \tilde{n_2}-\tilde{n_1}$. In other word, $N_k$ is defined as the minimal component number yielding $|\Delta (-1)|=3$ upon trial closing of the chain.The corresponding lateral size in axial direction $L_k$ characterizes the spatial size of the knot (Fig. \ref{fgr:scheme}(a)).
We note that the closure scheme generally has a finite probability to alter the topology at the stage of the closure using an supplemental arc, which results in the misjudging of the size and the type of the knot~\cite{MMO-rep}.
Fortunately, such an error rate can be made very low in our quasi-1D system by utilizing its geometrical property.

\section{Results}

As described above, the simulation was started from a prepared initial configuration with a trefoil knot in the middle of the chain. When the inserted knot size $N_k$ is large, we have observed that the system exhibits a spontaneous relaxation by expelling the strands toward the outside of the knotted part.
It then settles in a stationary state in which the knotted part diffuses along the chain keeping its average size.
A typical conformation in this stationary state is shown in Fig. \ref{fgr:scheme} (a) in the case of $D=6$ , $N=500$.

\begin{figure}[!t]
\centering
 \includegraphics[height=12cm]{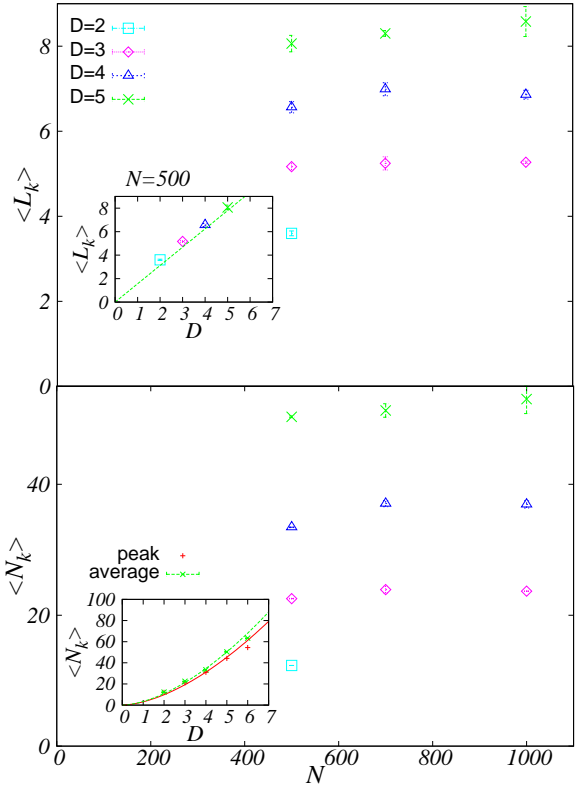}
  \caption{(Color online)Dependence of $\langle L_k \rangle$ and $\langle N_k \rangle$ for various tube sizes
on the total number of the monomers $N$.
(Inset)Dependence of $\langle L_k \rangle$, $\langle N_k \rangle$ and $N_k^{*}$ on $D$ for the chain length $N=500$.}
  \label{fgr:kitaichi}
\end{figure}

In Fig. \ref{fgr:kitaichi}, the averages of the knot size $\langle N_k \rangle$ and $\langle L_k \rangle$ are, respectively, plotted as functions of the total number of monomers $N$ for various tube size $D$.
Both $N_k$ and $L_k$ appear to be independent of the number of monomers of the polymer $N$, indicating the strong localization in 1D geometry.
\color{black}

\begin{figure}[]
  \centering
  \includegraphics[height=6.cm]{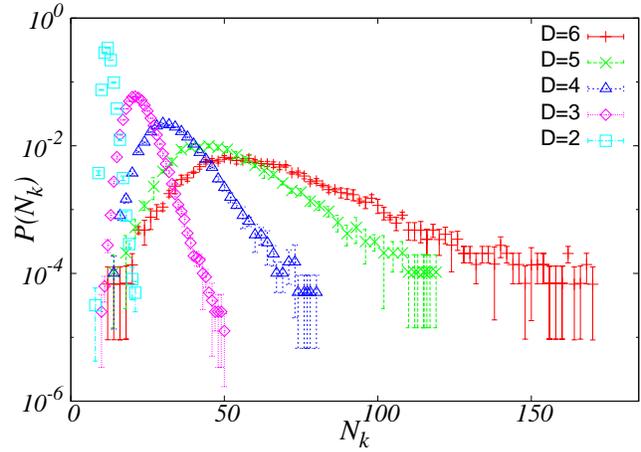}
  \caption{(Color online)Probability distributions of the knot size $N_k$ with various tube sizes.
}
  \label{fgr:bare}
\end{figure}

To investigate the behaviors of confined knotted polymers in more detail, we plot in Fig. ~\ref{fgr:bare} the probability distribution function $P(N_k)$ of the knot size in the stationary state for various tube sizes.
Here $P(N_k)$ is defined and obtained as follows.
We first build the histogram $H(N_k,L_k)$
where the pair values $(\hat{N_k},\hat{L_k})$ in each sample is located $N_k<\hat{N_k}<N_k+\Delta N_k$, and $L_k<\hat{L_k}<L_k+\Delta L_k$. Here we adopt $\Delta N_k=1$, $\Delta L_k=0.1$.
Then we approximate a two-dimensional density distribution of probability $P(N_k,L_k)$ as
\begin{eqnarray}
P(N_k,L_k) \simeq \frac{H(N_k,L_k)}{\big(\Delta N_k\big)\big(\Delta L_k\big)\big(\sum_{N_k',L_k'}H(N_k',L_k')\big)}.
\end{eqnarray}
From this density of probability, we construct the marginal distribution of $L_k$ at fixed $N_k$ as
\begin{eqnarray}
P(L_k | N_k) &\equiv& \frac{P(N_k,L_k)}{\int_{L_k}P(N_k,L_k)dL_k} \notag \\
&\simeq& \frac{H(N_k,L_k)}{\Delta L_k\sum_{N_k'}H(N_k',L_k')}
\end{eqnarray}
 and detect its peak position $L_k^{*}(N_k)$ for each value of $N_k$.
We then calculate $P(N_k)$ as $P(N_k,L_k^{*})$.
Because of the asymmetric profile of $P(N_k)$, the peak value $N_k^{*}$ of the knot size is slightly smaller than the average value $\langle N_k \rangle$,
but both exhibit qualitatively the same dependence on $D$ (see the inset of Fig. \ref{fgr:kitaichi}).
\begin{figure*}[]
\centering
\includegraphics[height=12.6cm]{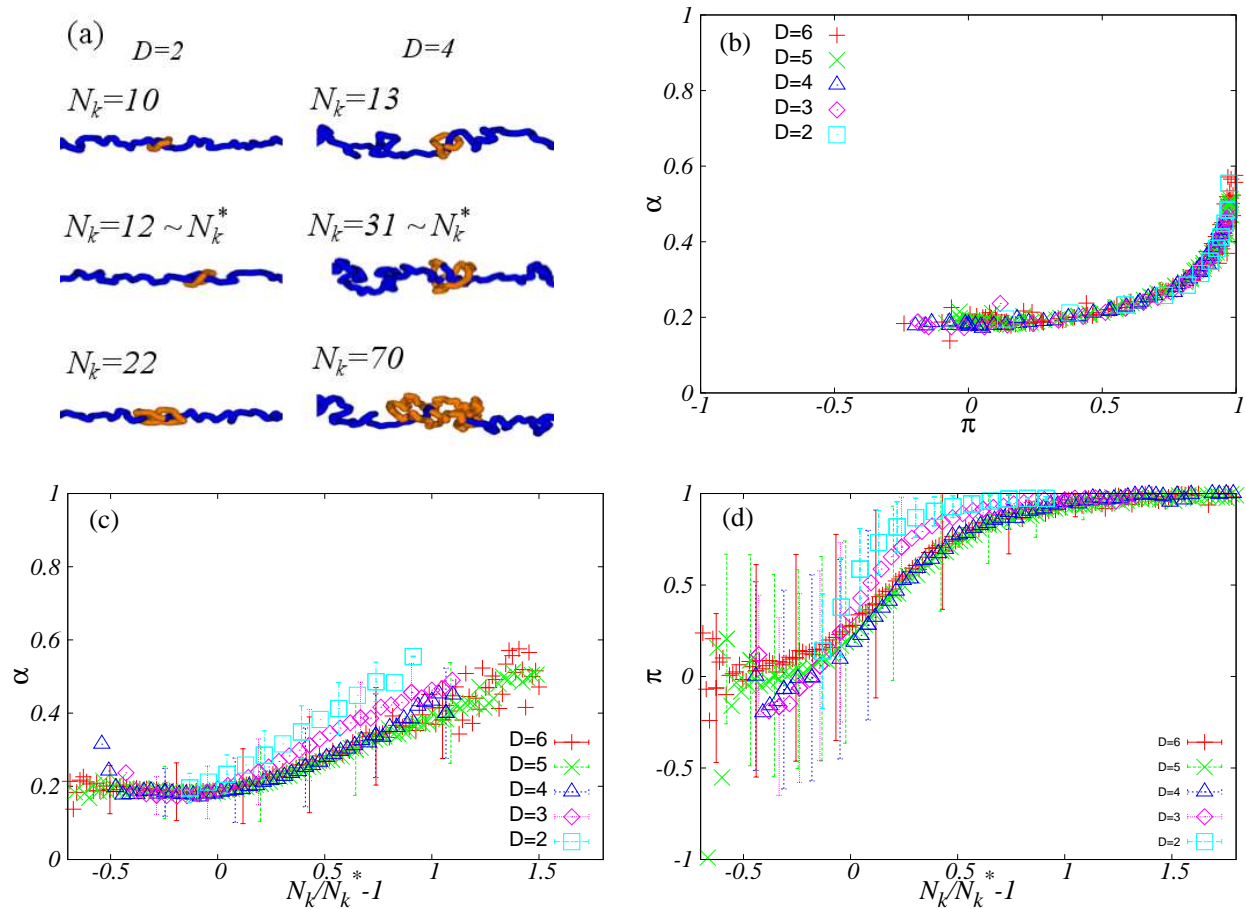}
  \caption{(Color online)(a)Typical confirmations of the knotted chain in the tube of size $D=2$ (left) and $D=4$ (right). Here shown are only the middle part of the whole chain containing the knotted part.
(b)Parametric plot of the relative asphericity  $\alpha$ and the relative prolateness $\pi$.
(c)Dependence of $\alpha$ as a function of the normalized knot size $N_k/N_k^{*}-1$.
(d)Dependence of $\pi$ as a function of the normalized knot size $N_k/N_k^{*}-1$.
Error bars in (c) and (d) indicates the standard deviation for rough measure of fluctuations.
}
  \label{fgr:shape_asph}
\end{figure*}
In Fig. \ref{fgr:shape_asph}(a), we show typical snapshots of the knotted polymer confined in tube with $D=2$(left) and $D=4$(right), where the size of the knotted part is small ($N_k<N_k^{*}$; top), in average ($N_k \simeq N_k^{*}$; center), large ($N_k>N_k^{*}$; bottom).

To quantitatively examine the spatial shape of the knotted part, we calculate the relative asphericity $\alpha$ and the relative prolateness $\pi$;
\begin{equation}
\alpha=\Bigg\langle \frac{\big(\lambda_1-\lambda_2\big)^2+\big(\lambda_2-\lambda_3\big)^2+\big(\lambda_3+\lambda_1\big)^2}{2\big(\lambda_1+\lambda_2+\lambda_3\big)^2}\Bigg\rangle,
\end{equation}
\begin{equation}
\pi=\Bigg\langle \frac{\big(2\lambda_1-\lambda_2-\lambda_3\big)\big(2\lambda_2-\lambda_3-\lambda_1\big)\big(2\lambda_3-\lambda_1-\lambda_2\big)}{2\big(\lambda_1^2+\lambda_2^2+\lambda_3^2+\lambda_1\lambda_2+\lambda_2\lambda_3+\lambda_3\lambda_1\big)^{3/2}} \Bigg\rangle
\end{equation}
where $\lambda_1 > \lambda_2 > \lambda_3$ are the eigenvalues of the gyration tensor of the knot.
As shown in Fig. \ref{fgr:shape_asph}(b), the ensemble of the knot shapes falls into a single curve in a $\alpha - \pi$ plane.
The shape asymptotically approaches the rod limit ($\alpha \rightarrow 1,\ \pi \rightarrow 1$) for large $N_k$.
The decrease in $N_k$ from the large $N_k$ limit is accompanied by the decrease in both $\alpha$ and $\pi$.
This trend continues up to the most probable point $N_k = N_k^*$, where $(\alpha,\pi)\simeq(0.19,0.) \Leftrightarrow \lambda_1 : \lambda_2 : \lambda_3 \simeq 2:1.5:1 $.
The value of $\alpha$ takes its minimum at $N_k^{*}$.
Further shrinkage of knot $N < N_k^{*}$ leads to the slight increase in $\alpha$, while $\pi$ remains almost constant $\pi=0$.
While in very narrow tubes ($D=2,$ and $3$), both $\alpha$ and $\pi$ take slightly larger values at a given normalized knot size $N_k/N_k^{*}-1$ [Fig. \ref{fgr:shape_asph}(c) and (d)], these seem to converge to asymptotic curves for larger tube sizes.
Note that in the region $N < N_k^{*}$, each value of the prolateness has large fluctuation and $N_k$ dependence of $\pi$ has large ambiguity. But these dependence seem to have a tendency to approach $\pi=0$ as $D$ becomes large.

\section{Discussion}\label{sec:overlapped}

\begin{figure}[]
\centering
\includegraphics[height=10.cm]{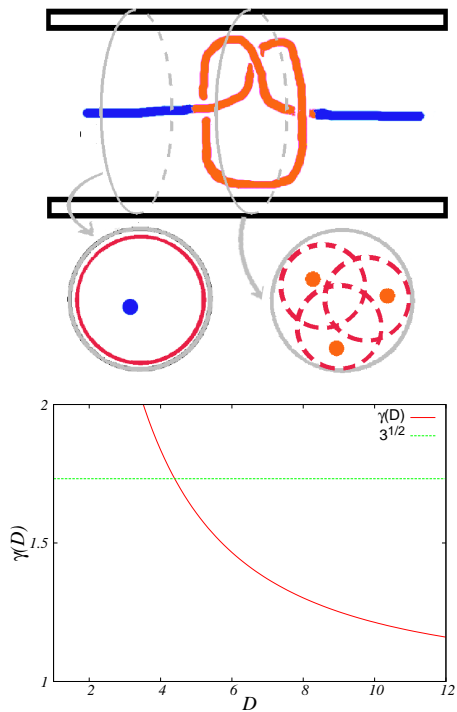}
  \caption{(Color online)(top)The schematic explanation of the concept of the virtual tube. 
(bottom)Functional form of the crowding factor $\gamma(D)$ as a function of $D$ with $A=150.$ and $B\simeq 2.9$.}
  \label{fgr:primitive_path}
\end{figure}

To understand the underlying mechanism of the strong knot localization and the behavior of the distribution function in Fig. \ref{fgr:bare}, we consider
the contribution of the free energy  for confining the knotted polymer into the tube.
The polymer can be divided into a knotted part with $N_k$ monomers in the middle and two ``arms" consisting of the $N-N_k$ remanent monomers at both sides.
Correspondingly we decompose the free energy $F$ into $F = F_k + F_A$.
The arms can be pictured as a linear sequence of blobs of size $D \simeq a g_A^{\nu}$, 
where $\nu \simeq 0.59$ is the Flory exponent and $g$ is the number of monomers per blob. The loss of degrees of freedom upon the confinement can be evaluated by assigning $\simeq k_BT$ per blob\cite{dG}. We thus have
\begin{eqnarray}\label{eq:f_e_a}
\frac{F_A}{k_BT} \simeq \frac{N-N_k}{g_A} \simeq \left(\frac{a}{D}\right)^{1/\nu} (N-N_k).
\end{eqnarray}

\begin{figure}[]
  \includegraphics[height=6.cm]{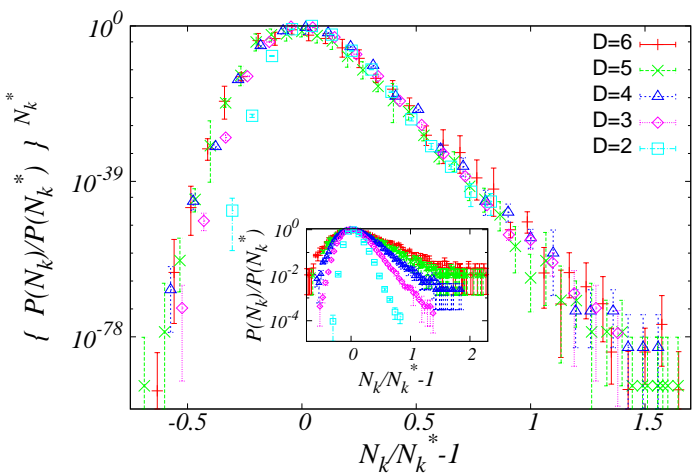}
  \caption{(Color online)Collapse of the probability distributions with the scaling variables $s=N_k/N_k^{*}-1$ and $\Big( P(N_k)/P(N_k^{*}) \Big)^{N_k^{*}}$, which corresponds to $q=1$ in Eq. \ref{eq:gen_scale}. The symbols are the same as those in Fig. \ref{fgr:bare}. The inset represents the failure of the collapse with $q=0$.}
  \label{fgr:scaled_conf}
\end{figure}

On the other hand, monomers in the knotted part feel more crowded environment.
If we look at a perpendicular cross-section ($X-Y$ plane) and count the number $p$ of ``primitive path" strands [see Fig. \ref{fgr:primitive_path}(top)], we find $p_A=1$ in arms and $p_k>p_A$ in the knotted part. In the present case of the trefoil knot in a polymer, we have $p_k=3$. 
The section of polymer in the knotted part is thus effectively confined to a narrower tube.
To implement such a feature, the simplest is to assume that each of $p_k$ strand in the knotted part is confined in the virtual tube whose cross-sectional area is $S_k=S/p_k$, where $S \simeq D^2$ represents the cross-sectional area of the original tube.
Then, the number of monomer $g_k$ per smaller blob due to the confinement is determined by the relation
\begin{eqnarray}
 \frac{D}{\gamma}\simeq a g_k^{\nu}
\end{eqnarray}
with the crowding factor $\gamma =p_k^{1/2} > 1$.
This leads to the confinement free energy of the knotted part as follows;
\begin{eqnarray}
\frac{F_k}{k_BT} \simeq \frac{N_k}{g_k} \simeq \gamma^{1/\nu}\left( \frac{a}{D}\right)^{1/\nu}N_k
\label{eq:f_K}
\end{eqnarray}
Using Eq. ~(\ref{eq:f_e_a}) and~(\ref{eq:f_K}), the total free energy of the confinement can be represented as
\begin{eqnarray}
\frac{F}{k_BT} \simeq \frac{F_0}{k_BT} + \left( \frac{a}{D}\right)^{1/\nu}N_k \ 
(\gamma^{1/\nu}-1)
\label{eq:f_total}
\end{eqnarray}
where $F_0 \simeq k_B TN(a/D)^{1/\nu}$ represents the conventional free energy of confining a linear polymer into tube, and the additional term is identified as the excess contribution due to the existence of the knot.
This argument is valid for relatively large size of knot, and indeed explains the exponential tail in the knot size probability distribution (Fig. \ref{fgr:bare} and \ref{fgr:scaled_conf} ) $P(N_k) \sim \exp{[-F(N_k)/k_BT]}$.
To reduce the crowded region as much as possible, the knot shrinks in 1D confined geometry.
The lower bound of $N_k$ for this shrinkage is provided by the tube size, below which the strands in the knotted part should be more highly confined with its topological constraint\cite{SR}.
We thus conclude that the knot size is determined by the tube size $\langle L_k \rangle \simeq D$.
Furthermore, if the optimum number of blobs in the knotted part $n/g_k$ has only weak or negligible dependence on the tube size $D$, we expect the relation $N_k \simeq \langle N_k \rangle \simeq (D/a)^{1/\nu}$.
These are consistent with the simulation results (Fig. \ref{fgr:bare} inset).

The virtual tube assumption seems to succeed to explain the qualitative trend including the strong knot localization and the exponential tail in the knot size probability distribution. 
We now examine it on a more quantitative basis by looking at the crowding factor $\gamma$.
To this end, we attempt to do a data collapse of the knot size distribution functions $P(N_k)$ for various tube sizes by seeking for the scaling function for $\Delta F = F(N_k)-F(N_k^*)$. According to Eq. ~(\ref{eq:f_total}), it is written as
\begin{eqnarray}
\frac{\Delta F(s)}{k_BT} \simeq B(\gamma^{1/\nu}-1)s
\label{m_eq}
\end{eqnarray}
where $s=(N_k/N_k^*)-1$ is the normalized deviation from the optimum size, and a numerical coefficient $B$ is introduced to express the relation between the optimum knot size and the tube size as $N_k^* = B(D/a)^{1/\nu}$.
Replotting the numerical results in the form of Eq. ~(\ref{m_eq}), it turned out that the naive assumption $\gamma = p_k^{1/2} = const.$ fails to collapse the data. Rather, we have found that the generalized form
\begin{eqnarray}\label{eq:gen_scale}
[N_k^{*}]^{q}\frac{\Delta F(s)}{k_BT} =A s
\label{m_eq_g}
\end{eqnarray}
with an exponent $q\simeq 1$ provides a very good data collapse  (See Fig. \ref{fgr:scaled_conf}).
The value of $A$ is roughly estimated as $A \simeq 120. \sim 150.$ from the slope of the exponential tails of Fig. \ref{fgr:scaled_conf}.
Comparing Eqs.~(\ref{m_eq}) and~(\ref{m_eq_g}), we find that the crowding factor should depend on the tube size as
\begin{eqnarray}\label{eq:overlap_argument}
\gamma (D) = \left[ 1+ \frac{A}{B^{1+q}}\left( \frac{a}{D}\right)^{q/\nu}\right]^{\nu}
\end{eqnarray}
As shown in the bottom of Fig. \ref{fgr:primitive_path}, the crowding factor $\gamma$ decreases with the increase in the tube size.
This result could be interpretted as follows; while the most outer part of the assumed virtual tubes are composed of the original rather rigid wall potential, their inner parts originate from softer fluctuating potential field due to the presence of other primitive path strands.
This allows the partial overlap of the virtual tubes, the degree of which may increase with the tube size.
The narrower the tube size is, the more the fluctuation of the primitive strands is suppressed.
The $D$ dependence of $N_k^{*}$ shown in the inset of Fig. \ref{fgr:kitaichi} determines $B\simeq 2.9$.
Using these independently determined value $A$ and $B$ in Eq. \ref{eq:overlap_argument}, $D_0 \simeq 4.$ is estimated as the value at which $\gamma(D_0)=p_k^{1/2}$ with $p_k=3$.
The tube size $D_0$ signifies the very narrow space, inside which $p_k$ strands tightly fit to fill it.
In fact, the slight deviations of the data points with $D=2$ and $3$ from the master curve in Fig. \ref{fgr:shape_asph} and \ref{fgr:scaled_conf} are probably attributed to it.
Although based on a rather rough estimation, such an internal consistency is regarded as a support for the line of our argument.

\section{Summary and perspectives}
 There have been several recent studies on the (de)localization property of knots formed in a ring polymer.
It has been argued that, for flat knots (2D), the entropy of the looping is a relevant factor lending to the strong knot localization, the essential feature of which can be understood using the slip-link model\cite{MHDKK}.
In the present paper, we have shown that the knot is strongly localized in tubes (1D), too, but its physical mechanism is entirely different from that in 2D case.
As might be evident from the analysis of the slip-link model, what is relevant here is not the entropy of the looping but the crowding due to the presence of parallel strands in the narrow tube.

A useful way of taking such a crowding effect into account is to introduce the virtual tube concept, in which a plural number $p_k$ of tubes are supposed to be present in accordance with the number $p_k$ of parallel strands.
This naturally leads to the exponential probability distribution for the larger side of knot size, which has been well confirmed by numerical simulations.

Our closer analysis of the knot size probability distribution has revealed that the crowding factor generally depends not only on the number $p_k$ of the parallel strands but also on the size of confining tubes.
In addition to the knot localization phenomena, we expect that the present result would have an important consequence on the behaviors of circular polymer ($p_k=2$) 
which would be relevant to the bacterial chromosome segregation\cite{Jun} as well as the migration dynamics in nano-channels \cite{MRFK}.
Another related system is a star polymer with $f$ arms confined in tube, where we may set $p_k=f_l,f_r$ with $f_l$, $f_r=f-f_l$ being the number of arms in left and right sides of the center\cite{BdG}. 

Our main conclusion that the knot is strongly localized in 1D geometry is in apparent contradiction with the prediction by M\"{o}bius et al. \cite{MFG}.
While the influence of the chain semiflexibility included in the ref [12] still remains to be clarified, we expect that the presence of the crowding factor would be robust in 1D geometries.
This leads us to propose that the predominant mode of the unknotting dynamics in 1D geometries is not a knot inflation, but a diffusion of the localized knot along the chain.

\section*{Acknowledgement}
 This work was supported by the JSPS Core-to-Core Program ``International research network for non-equilibrium dynamics of soft matter''.

\renewcommand\refname{Notes and references}

\footnotesize{

}

\end{document}